\documentclass[conference]{IEEEtran}
\IEEEoverridecommandlockouts
\usepackage{cite}
\usepackage{amsmath,amssymb,amsfonts}
\usepackage{graphicx}
\usepackage{textcomp}
\usepackage{xcolor}
\def\BibTeX{{\rm B\kern-.05em{\sc i\kern-.025em b}\kern-.08em
    T\kern-.1667em\lower.7ex\hbox{E}\kern-.125emX}}

\usepackage{makecell}

\usepackage{threeparttable}

\usepackage{color,soul}
\soulregister\cite7

\usepackage{algorithm}
\usepackage{algpseudocode}

\usepackage{amsthm}

\newtheorem{theorem}{Theorem}
\newtheorem{proposition}{Proposition}
 
\newtheorem{lemma}{Lemma}
\newtheorem{assumption}{Assumption}

\newtheorem{remark}{Remark}

\begin{document}

\title{\huge Global Optimization of Energy Efficiency in IRS-Aided \\ Communication Systems via Robust IRS-Element Activation  \\
\thanks{This work was supported by the Research Promotion Foundation, Cyprus, under the project INFRASTRUCTURES/1216/0017 (IRIDA). This work was also supported by the European Research Council (ERC) under the European
Union's Horizon 2020 research and innovation programme (Grant agreement No. 819819). \newline  
\indent This article has been accepted for publication in \textit{IEEE International Conference on Communications (ICC), 2023}. Copyright \copyright 2023 IEEE. Personal use is permitted, but republication/redistribution requires IEEE permission.} 
}

\author{\IEEEauthorblockN{Christos N. Efrem and Ioannis Krikidis}
\IEEEauthorblockA{Department of Electrical and Computer
Engineering  \\
University of Cyprus, Nicosia, Cyprus \\
\{efrem.christos, krikidis\}@ucy.ac.cy}
}

\maketitle

\begin{abstract}
In this paper, we study an intelligent reflecting surface (IRS) assisted communication system with single-antenna transmitter and receiver, under imperfect channel state information (CSI). More specifically, we deal with the robust selection of binary (on/off) states of the IRS elements in order to maximize the worst-case energy efficiency (EE), given a bounded CSI uncertainty, while satisfying a minimum signal-to-noise ratio (SNR). The IRS phase shifts are adjusted so as to maximize the ideal SNR (i.e., without CSI error), based only on the estimated channels. First, we derive a closed-form expression of the worst-case SNR, and then formulate the robust (discrete) optimization problem. Moreover, we design and analyze a dynamic programming (DP) algorithm that is theoretically guaranteed to achieve the global maximum with polynomial complexity $O(L\,{\log L})$, where $L$ is the number of IRS elements. Finally, numerical simulations confirm the theoretical results. In particular, the proposed algorithm shows identical performance with the exhaustive search, and significantly outperforms a baseline scheme, namely, the activation of all IRS elements. 

\end{abstract}


\section{Introduction}

Intelligent reflecting surfaces (IRSs) have emerged as a promising technology for the dynamic configuration of electromagnetic waves \cite{Liaskos2018,DiRenzo2019}. A key characteristic of passive IRSs is low energy consumption. Energy efficiency (EE) optimization has been an attractive area of research in wireless networks \cite{Efrem2019a}, and more recently in IRS-aided communication systems \cite{Huang2019}. Moreover, there has been a growing interest in active/passive-IRS beamforming to optimize the achievable sum-rate \cite{Zhao2021}, signal-to-noise ratio (SNR) \cite{Long2021,Zhang2022}, and transmit power \cite{Zhou2020}. 

In addition, the joint optimization of IRS location and size (number of IRS elements) to minimize the outage probability has been studied in \cite{Efrem2022}. By considering channel estimation and feedback, the number of IRS elements maximizing spectral efficiency (SE), EE, and SE-EE tradeoff has been investigated in \cite{Zappone2021}. Furthermore, the author in \cite{Li2022} has examined the minimum number of IRS elements satisfying minimum-SE/EE requirements. Last but not least, a low-cost scheme for optimizing only the on/off states of the IRS elements, while fixing their phase shifts, has been proposed in \cite{Khaleel2022}.

The main contributions of this paper are the following: 

\begin{itemize}
\item First of all, we provide a closed-form expression of the worst-case SNR under channel state information (CSI) uncertainty, which is then used to formulate a robust optimization problem. In particular, we aim to maximize the worst-case EE, subject to a minimum-SNR constraint, by tuning the on/off states of IRS elements. The IRS phase shifts are selected so as to maximize the ideal SNR, based merely on the estimated channels. On the other hand, existing works dealing with the activation of IRS elements assume perfect CSI or only phase errors \cite{Zappone2021,Khaleel2022}, fixed IRS phase shifts \cite{Khaleel2022}, optimize a different performance metric other than EE \cite{Zhao2021,Li2022,Khaleel2022}, and do not consider a minimum-SNR constraint \cite{Zappone2021}. 

\item In addition, we develop a polynomial-time dynamic programming (DP) algorithm with global optimization guarantee. To the best of our knowledge, this is the first work that achieves global optimization of EE (with polynomial complexity) in IRS-aided communication systems by taking into account imperfect CSI knowledge. 

\item Furthermore, the algorithm performance is validated through numerical results. Specifically, the DP algorithm and exhaustive search both attain a (globally) optimal solution, while outperforming a conventional scheme.    

\end{itemize}

The rest of this paper is organized as follows. Section \ref{section:System_Model} describes the system model, Section \ref{section:Problem_Formulation} formulates the robust optimization problem, and Section \ref{section:Dynamic_Programming} develops and analyzes the proposed algorithm. Moreover, Section \ref{section:Numerical_Results} presents some numerical simulations, while Section \ref{section:Conclusion} concludes the paper.

\section{System Model} \label{section:System_Model}

\subsection{Signal Transmission Model}

Consider a wireless system consisting of a single-antenna transmitter (Tx) and a single-antenna receiver (Rx) that communicate through a passive IRS with $L$ elements.\footnote{An IRS is called ``passive'' if all its elements are passive (i.e., their induced amplitude is less than or equal to 1). On the other hand, it is called ``active'' if some of its elements are active (i.e., their amplitude is greater than 1).} Without loss of generality, the set of all IRS elements is denoted by $\mathcal{L} = \{1,\dots,L\}$. Let $h_0 \in \mathbb{C}$ be the channel coefficient of the Tx-Rx direct link, and ${{\mathbf{u}}} = {[{u_1}, \ldots ,{u_L}]^ \top } \in \mathbb{C}^L$, ${{\mathbf{v}}} = {[{v_1}, \ldots ,{v_L}]^ \top \in \mathbb{C}^L}$ be the channel-coefficient vectors of Tx-IRS and IRS-Rx links, respectively. All channel coefficients are constant within a given time-slot (flat-fading channels). 

Moreover, each IRS element can be either on (i.e., activated) or off (i.e., deactivated). In particular, we say that an IRS element is \emph{``on''} when operating in standard reflection mode (\emph{reflecting element}), whereas it is \emph{``off''} when functioning in absorption mode (\emph{absorbing element}) \cite{Zhao2021,Khaleel2022}. For this purpose, we introduce a binary vector ${{\mathbf{x}}} = {[{x_1}, \ldots ,{x_L}]^ \top \in \{0,1\}^L}$, so that $x_\ell = 1$ if and only if the $\ell^\text{th}$ IRS element is on; equivalently, $x_\ell = 0$ if and only if the $\ell^\text{th}$ IRS element is off. The IRS on/off-state-and-phase-shift diagonal matrix is given by ${{\mathbf{D}}} (\mathbf{x},\boldsymbol{\phi}) = {\operatorname{diag}}({x_1 \beta_1 e^{j{\phi_1}}}, \ldots ,{x_L \beta_L e^{j{\phi _{L}}}})$, where $\beta_\ell \in [0,1]$ is the amplitude attenuation and $\phi_\ell \in [0,2\pi)$ is the phase shift of the $\ell^{\text{th}}$ IRS element (with ${\boldsymbol{\phi}} = [{\phi_1}, \ldots ,{\phi_L}]^\top \in [0,2\pi)^L$). In this work, we assume that the amplitude coefficients $\{ \beta_\ell \}_{\ell \in \mathcal{L}}$ are fixed, and the bit-resolution of phase shifts is infinite (continuous phase shifts).\footnote{In practice, if the number of bits used for controlling the IRS phase shifts is sufficiently large, then the quantization error becomes negligible.}  

We also define the cascaded channel coefficient corresponding to the $\ell^\text{th}$ IRS element, i.e., $h_\ell = {\beta_\ell} {u_\ell} {v_\ell} \in \mathbb{C}$, for all $\ell \in \mathcal{L}$. The complex channel coefficients (direct and cascaded) are expressed in polar coordinates, i.e., ${h_\ell} = {\alpha_\ell} {e^{j\theta_\ell}}$, where $\alpha_\ell = | h_\ell | \geq 0$ is the amplitude and $\theta_\ell = \arg(h_\ell) \in [0,2\pi)$ is the principal argument of $h_\ell$, for all $\ell \in \mathcal{L}_0 = \{0,\dots,L\}$. For convenience, the direct and cascaded channel coefficients are grouped in vector form as $\mathbf{h} = [h_0,\dots,h_L]^\top \in \mathbb{C}^{L+1}$.

The received signal is a superposition of the transmitted signal through the direct and cascaded channels, i.e.,
\begin{equation}
\begin{split}
r & = \sqrt{p} \left(h_0 + \mathbf{u}^\top {\mathbf{D}}(\mathbf{x},\boldsymbol{\phi}) \, \mathbf{v} \right) s + w  \\
& = \sqrt{p} \left( h_0 + \sum_{\ell \in \mathcal{L}} {x_\ell h_\ell  e^{j\phi_\ell} } \right)s + w   ,
\end{split}  
\end{equation}
where $p > 0$ is the (fixed) transmit power, $s \in \mathbb{C}$ is the transmitted symbol (with $\mathbb{E}(|s|^2)=1$), and $w \sim \mathcal{CN}(0,\sigma_w^2)$ is the additive white Gaussian noise (AWGN) at the receiver (with ${\sigma_w^2}>0$ being the noise power). Therefore, the SNR at the receiver is given by 
\begin{equation} \label{equation:SNR_initial}
\gamma = \overline{\gamma} \left|  h_0 + \sum_{\ell \in \mathcal{L}} {x_\ell h_\ell  e^{j\phi_\ell} }  \right| ^2 ,
\end{equation}
where $\overline{\gamma} = p/{\sigma_w^2} > 0$.

\subsection{Imperfect CSI Model and Assumptions}

Regarding the CSI acquisition, we assume that the receiver estimates the direct and cascaded channel coefficients, and then feeds the (vector) quantized CSI back to the transmitter. As a result, the obtained CSI is subject to quantization errors.

Herein, we adopt an additive CSI-error model as follows
\begin{equation} \label{equation:CSI-error_model}
\mathbf{h} = \widehat{\mathbf{h}} + \widetilde{\mathbf{h}}  ,
\end{equation}
where $\widehat{\mathbf{h}} = [\widehat{h}_0,\dots,\widehat{h}_L]^\top \in \mathbb{C}^{L+1}$ is the estimated channel vector (\emph{known} at the transmitter), while $\widetilde{\mathbf{h}} = [\widetilde{h}_0,\dots,\widetilde{h}_L]^\top \in \mathbb{C}^{L+1}$ is the \emph{unknown} channel estimation error. As before, we can express these complex coefficients in polar coordinates, i.e., ${\widehat{h}_\ell} = {\widehat{\alpha}_\ell} {e^{j\widehat{\theta}_\ell}}$ and ${\widetilde{h}_\ell} = {\widetilde{\alpha}_\ell} {e^{j\widetilde{\theta}_\ell}}$, for all $\ell \in \mathcal{L}_0$.

In order to avoid making any assumptions on the statistics of $\widetilde{\mathbf{h}}$, we consider a (deterministic) \emph{bounded CSI-error model}. More specifically, we assume that $\widetilde{\mathbf{h}}$ lies within a compact (closed and bounded) ball of uncertainty, i.e.,
\begin{equation} \label{equation:CSI_uncertainty}
{\lVert \widetilde{\mathbf{h}} \rVert}_2 \triangleq \sqrt{ \sum_{\ell \in \mathcal{L}_0} {| \widetilde{h}_\ell |^2} } = \sqrt{ \sum_{\ell \in \mathcal{L}_0} {\widetilde{\alpha}_\ell^2} } \leq \delta  ,
\end{equation}
where ${\lVert \cdot \rVert}_2$ denotes the Euclidean norm, and $\delta \geq 0$ is the radius of the CSI-uncertainty region \emph{known} at the transmitter \cite{Zhou2020,Shenouda2007}. Note that by setting $\delta = 0$ we obtain the case of \emph{perfect CSI} (i.e., without channel estimation errors), since $\delta = 0$ implies $\widetilde{\mathbf{h}} = \mathbf{0}_{L+1}$ and therefore $\widehat{\mathbf{h}} = \mathbf{h}$.

Furthermore, we make the following assumption that will be useful in Section \ref{section:Problem_Formulation}. 

\begin{assumption} \label{assumption:CSI-uncertainty_radius_1}
The CSI-uncertainty radius, $\delta$, satisfies the following inequality 
\begin{equation} \label{equation:delta_psi}
\delta \leq  \min_{M \in \mathcal{L}_0} \left\{ \frac{\widehat{\alpha}_0 + \sum_{m=1}^{M} {\widehat{\alpha}_{\sigma_m}}}{\sqrt{1 + M}} \right\}  \triangleq  \psi ,
\end{equation}
where $(\sigma_1,\dots,\sigma_L)$ is a permutation of $\mathcal{L}$ such that $\widehat{\alpha}_{\sigma_1} \leq \cdots \leq \widehat{\alpha}_{\sigma_L}$. Equivalently, $\widehat{\alpha}_0 + \sum_{m=1}^{M} {\widehat{\alpha}_{\sigma_m}} \geq \delta \sqrt{1 + M}$, $\forall M \in \mathcal{L}_0$.
\end{assumption}

\noindent Finally, we provide another assumption that is stronger, but simpler, than Assumption \ref{assumption:CSI-uncertainty_radius_1}. 

\begin{assumption} \label{assumption:CSI-uncertainty_radius_2}
The CSI-uncertainty radius, $\delta$, is upper bounded by the minimum amplitude of estimated channel coefficients, i.e., $\delta \leq \min_{\ell \in \mathcal{L}_0} \{ \widehat{\alpha}_\ell \} \triangleq \widehat{\alpha}_{\min}$. 
\end{assumption}

\begin{proposition} \label{proposition:Assumptions_1_2}
Assumption \ref{assumption:CSI-uncertainty_radius_2} implies Assumption \ref{assumption:CSI-uncertainty_radius_1}.  
\end{proposition}

\begin{IEEEproof}
Suppose that Assumption \ref{assumption:CSI-uncertainty_radius_2} is true. Then, $\psi \geq \min_{M \in \mathcal{L}_0} \left\{ \frac{\widehat{\alpha}_{\min} \left(1 + M \right)}{\sqrt{1 + M}} \right\} = \widehat{\alpha}_{\min} \min_{M \in \mathcal{L}_0} \left\{ \sqrt{1+M} \right\} = \widehat{\alpha}_{\min} \geq \delta$, thus Assumption \ref{assumption:CSI-uncertainty_radius_1} is true as well. 
\end{IEEEproof}
\vspace{0.5mm}

\subsection{IRS Phase-Shift Design}

Under the CSI-error model given by \eqref{equation:CSI-error_model}, the SNR in \eqref{equation:SNR_initial} can be written as 
\begin{equation}
\gamma (\mathbf{x},\widetilde{\mathbf{h}},\boldsymbol{\phi}) = \overline{\gamma} \left|  \widehat{h}_0 + \sum_{\ell \in \mathcal{L}} {x_\ell \widehat{h}_\ell e^{j\phi_\ell} } + \widetilde{h}_0 + \sum_{\ell \in \mathcal{L}} {x_\ell \widetilde{h}_\ell e^{j\phi_\ell} } \right| ^2 .
\end{equation}
Since the channel estimation error, $\widetilde{\mathbf{h}}$, is unknown, we select the IRS phase shifts so as to maximize the ideal SNR without CSI error ($\widetilde{\mathbf{h}} = \mathbf{0}_{L+1}$). In particular, the optimal phase shifts are given by\footnote{Here, we use the modulo operation in order to ensure that $\phi_\ell^\star \in [0,2\pi)$. More precisely, for any $y \in \mathbb{R}$ and any $a > 0$, the modulo operation is defined as $y \bmod a = y - a \lfloor y/a \rfloor$, where $\lfloor \cdot \rfloor$ is the floor function. Note that $y \bmod {a} \in [0,a)$, because $z - 1 < \lfloor z \rfloor \leq z$, for all $z \in \mathbb{R}$.}
\begin{equation}
\phi_\ell^{\star} =  (\widehat{\theta}_0 - \widehat{\theta}_\ell) \bmod {2\pi}, \;\; \forall \ell \in \mathcal{L} . 
\end{equation} 
The SNR achieved by the optimal phase shifts is expressed as
\begin{equation} \label{equation:SNR_optimal_phase_shifts}
\begin{split}
{\gamma} (\mathbf{x},\widetilde{\mathbf{h}}) & \triangleq \gamma (\mathbf{x},\widetilde{\mathbf{h}},{\boldsymbol{\phi}}^{\star}) \\
& = \overline{\gamma} \left| f(\mathbf{x}) e^{j\widehat{\theta}_0} + \widetilde{h}_0 + \sum_{\ell \in \mathcal{L}} {x_\ell \widetilde{h}_\ell e^{j{\phi}_\ell^{\star}}} \right| ^2  ,
\end{split}
\end{equation}
where  
\begin{equation} \label{equation:f}
f(\mathbf{x}) =  \widehat{\alpha}_0 + \sum_{\ell \in \mathcal{L}} {x_\ell \widehat{\alpha}_\ell}  .
\end{equation}

\subsection{Power Consumption Model}

The total power consumption is computed as follows
\begin{equation}
\begin{split}
\operatorname{P}_\text{tot} (\mathbf{x}) & = P_{\text{fix}} + P_{\text{on}} {L_{\text{on}} (\mathbf{x})} + P_{\text{off}} {L_{\text{off}} (\mathbf{x})}  \\
& =  P_{\text{fix}} + L P_{\text{off}} + (P_{\text{on}} - P_{\text{off}}) {\sum_{\ell \in \mathcal{L}} {x_\ell}}  ,
\end{split}
\end{equation}
where $P_{\text{fix}} = p/\eta + P_{\text{static}}$, with $\eta \in (0,1]$ being the efficiency of transmitter's power amplifier, and $P_{\text{static}} > 0$ accounting for the dissipated power in the remaining signal processing blocks at the transmitter and receiver. In addition, $L_{\text{on}} (\mathbf{x}) = \sum_{\ell \in \mathcal{L}} {x_\ell}$ and $L_{\text{off}} (\mathbf{x}) = L - L_{\text{on}} (\mathbf{x})$ is the number of activated and deactivated IRS elements, respectively. Furthermore, $P_{\text{on}}$, $P_{\text{off}} > 0$ represent the power consumption of each activated/deactivated element, respectively. Deactivated IRS elements usually consume a relatively small (albeit, non-negligible) amount of power compared to that of activated elements, with $P_{\text{off}} \leq P_{\text{on}}$. Similar power consumption models for passive/active IRSs can be found in \cite{Long2021}.

\section{Formulation of Robust Optimization Problem} \label{section:Problem_Formulation}

The worst-case SNR is defined as the lowest SNR that can be attained under the CSI uncertainty given by \eqref{equation:CSI_uncertainty}, i.e.,
\begin{subequations}  \label{problem:Worst-case_SNR}
\begin{alignat}{3}
  {\gamma}_{\text{worst}} (\mathbf{x};\delta) \! \triangleq & \mathop {\text{minimize}} \limits_{\widetilde{\mathbf{h}} \in \mathbb{C}^{L+1}} & \quad & {\gamma} (\mathbf{x},\widetilde{\mathbf{h}})  \\
& \,\text{subject to} & & {\lVert \widetilde{\mathbf{h}} \rVert}_2  \leq  \delta . \label{constraint:CSI_uncertainty}
\end{alignat}
\end{subequations}
The global minimum of problem \eqref{problem:Worst-case_SNR} admits a closed-form expression, according to the following theorem.

\begin{theorem} \label{theorem:Worst-case_SNR}
Suppose that Assumption \ref{assumption:CSI-uncertainty_radius_1} holds. Then, an optimal solution to problem \eqref{problem:Worst-case_SNR} is given by
\begin{subequations}  
\begin{gather}
  \widetilde{\theta}_0^\star = (\widehat{\theta}_0 + \pi) \bmod {2\pi}  , \\
 \widetilde{\theta}_\ell^\star = (\widehat{\theta}_0 - {\phi}_\ell^{\star} + \pi) \bmod {2\pi} ,  \;\;\forall \ell \in \mathcal{L} , \\
  \widetilde{\alpha}_0^\star = \widetilde{\alpha}_\ell^\star = \lambda ,  \;\;\forall \ell \in \mathcal{P} , \\
 \widetilde{\alpha}_\ell^\star = 0 ,  \;\;\forall \ell \in {\mathcal{L} \setminus \mathcal{P}} ,
\end{gather}
\end{subequations}
where $\lambda = \delta \big/ \sqrt{1 + \sum_{l \in \mathcal{L}} {x_l}}$ and $\mathcal{P} = \{\ell \in \mathcal{L} : \, x_\ell = 1 \}$. In addition, the worst-case SNR is expressed in closed form as  
\begin{equation}  \label{equation:Worst-case_SNR} 
{\gamma}_{\textnormal{worst}} (\mathbf{x};\delta) = \overline{\gamma} \left( f(\mathbf{x}) - g(\mathbf{x};\delta) \right)^2,
\end{equation}
where $f(\mathbf{x})$ is given by \eqref{equation:f}, and 
\begin{equation} \label{equation:g}
g(\mathbf{x};\delta) = \delta \sqrt{1 + \sum_{\ell \in \mathcal{L}} {x_\ell}} \, .
\end{equation}
Finally, $f(\mathbf{x}) \geq g(\mathbf{x};\delta), \ \forall \mathbf{x} \in \{0,1\}^L$.
\end{theorem}

\begin{IEEEproof}
See Appendix \ref{appendix:A}.
\end{IEEEproof}

\begin{remark} \label{remark:Assumption_replacement}
Theorem \ref{theorem:Worst-case_SNR} is still valid if Assumption \ref{assumption:CSI-uncertainty_radius_1} is replaced by Assumption \ref{assumption:CSI-uncertainty_radius_2}, because of Proposition \ref{proposition:Assumptions_1_2}.
\end{remark}

\noindent Based on \eqref{equation:Worst-case_SNR}, the ideal worst-case SNR (i.e., without CSI error) is equal to ${\gamma}_{\text{worst}}^\text{ideal} (\mathbf{x}) = {\gamma}_{\text{worst}} (\mathbf{x};0) = \overline{\gamma} (f(\mathbf{x}))^2$. 

Subsequently, the worst-case spectral and energy efficiency are respectively expressed as 
\begin{equation}
{\operatorname{SE}_{\text{worst}}}(\mathbf{x};\delta) = \log_2 \left( 1 + {\gamma_{\text{worst}}} (\mathbf{x};\delta) \right)  ,
\end{equation}
\begin{equation}
{\operatorname{EE}_{\text{worst}}}(\mathbf{x};\delta) = \frac{{\operatorname{SE}_{\text{worst}}}(\mathbf{x};\delta)}{\operatorname{P}_\text{tot} (\mathbf{x})}  .
\end{equation}

\noindent The \emph{robust (discrete) optimization problem} is formulated as 
\begin{subequations} \label{problem:EE_original}
\begin{alignat}{3}
 \operatorname{EE}_{\text{worst}}^\star (\delta,\gamma_{\min}) \! \triangleq & \mathop {\text{maximize}} \limits_{\mathbf{x} \in \{ 0,1 \}^L} & \quad & {\operatorname{EE}_{\text{worst}}}(\mathbf{x};\delta)   \\
  & \,\text{subject to} & & {\gamma_{\text{worst}}} (\mathbf{x};\delta) \geq \gamma_{\min} , 
\end{alignat}
\end{subequations}
where $\gamma_{\min} \geq 0$ is the minimum required SNR. In other words, we are looking for the optimal on/off states of IRS elements in order to maximize the worst-case EE, given the CSI uncertainty, while satisfying a minimum-SNR constraint. 

A necessary and sufficient condition for feasibility as well as monotonicity properties of ${\gamma_{\text{worst}}} (\mathbf{x};\delta)$ and ${\operatorname{EE}_{\text{worst}}}(\mathbf{x};\delta)$ are provided by the following propositions. 

\begin{proposition} \label{proposition:Feasibility}
Under Assumption \ref{assumption:CSI-uncertainty_radius_2} (or Assumption \ref{assumption:CSI-uncertainty_radius_1} and $\widehat{\alpha}_{\sigma_1} \triangleq \min_{\ell \in \mathcal{L}} \{ \widehat{\alpha}_\ell \} \geq \widehat{\alpha}_0$), ${\gamma_{\textnormal{worst}}} (\mathbf{x};\delta)$ is nondecreasing in each variable, i.e., ${\partial {\gamma_{\textnormal{worst}}} (\mathbf{x};\delta)} / {\partial x_\ell} \geq 0$, $\forall \ell \in \mathcal{L}$. Therefore, the discrete optimization problem \eqref{problem:EE_original} is feasible if and only if ${\gamma}_{\textnormal{worst}} (\mathbf{1}_L;\delta) \geq \gamma_{\min}$.
\end{proposition}

\begin{IEEEproof}
First, Assumption \ref{assumption:CSI-uncertainty_radius_1} implies $\widehat{\alpha}_0 \geq \delta$ and together with $\widehat{\alpha}_{\sigma_1} \geq \widehat{\alpha}_0$ yield $\widehat{\alpha}_{\min} = \min\{\widehat{\alpha}_0,\widehat{\alpha}_{\sigma_1}\} = \widehat{\alpha}_0 \geq \delta$, which gives Assumption \ref{assumption:CSI-uncertainty_radius_2}. Next, provided that Assumption \ref{assumption:CSI-uncertainty_radius_2} is true, Theorem \ref{theorem:Worst-case_SNR} is applicable due to Remark \ref{remark:Assumption_replacement}. The partial derivative of \eqref{equation:Worst-case_SNR}, with respect to an arbitrary variable $x_\ell$, is expressed as ${\partial {\gamma_{\text{worst}}} (\mathbf{x};\delta)} / {\partial x_\ell} = \overline{\gamma} (f(\mathbf{x}) - g(\mathbf{x};\delta)) \left( 2\widehat{\alpha}_\ell - \delta \big/ \sqrt{1 + \sum_{\ell \in \mathcal{L}} {x_\ell}} \right)$ with $\overline{\gamma} \geq 0$, $f(\mathbf{x}) - g(\mathbf{x};\delta) \geq 0$, and $2\widehat{\alpha}_\ell - \delta \big/ \sqrt{1 + \sum_{\ell \in \mathcal{L}} {x_\ell}}  \geq  2\widehat{\alpha}_{\min} - \delta \geq 2\delta - \delta = \delta \geq 0$. Hence, ${\partial {\gamma_{\text{worst}}} (\mathbf{x};\delta)} / {\partial x_\ell} \geq 0$. But $x_\ell$ is arbitrarily chosen, thus ${\partial {\gamma_{\text{worst}}} (\mathbf{x};\delta)} / {\partial x_\ell} \geq 0$, $\forall \ell \in \mathcal{L}$. As a result, ${\gamma}_{\textnormal{worst}} (\mathbf{1}_L;\delta) \geq {\gamma_{\text{worst}}} (\mathbf{x};\delta)$, $\forall \mathbf{x} \in \{0,1\}^L$, which easily leads to the if-and-only-if statement. 
\end{IEEEproof}

\begin{proposition} \label{proposition:Optimal_worst-case_EE_monotonicity}
Under Assumption \ref{assumption:CSI-uncertainty_radius_1}, ${\gamma_{\textnormal{worst}}} (\mathbf{x};\delta)$ and ${\operatorname{EE}_{\textnormal{worst}}}(\mathbf{x};\delta)$ are nonincreasing functions of $\delta$, i.e., ${\partial {\gamma_{\textnormal{worst}}} (\mathbf{x};\delta)} / {\partial \delta} \leq 0$ and ${\partial {\operatorname{EE}_{\textnormal{worst}}}(\mathbf{x};\delta)} / {\partial \delta} \leq 0$. Furthermore, the optimal worst-case EE, defined by \eqref{problem:EE_original}, is nonincreasing in each of its arguments, i.e., 
\begin{equation}  
\begin{split}
  \delta_1 \leq \delta_2 \implies & \operatorname{EE}_{\textnormal{worst}}^\star (\delta_1,\gamma_{\min}) \geq \operatorname{EE}_{\textnormal{worst}}^\star (\delta_2,\gamma_{\min}) , \\
  & \; \forall \gamma_{\min} \geq 0  , 
\end{split}  
\end{equation}
\begin{equation}
\begin{split}
  \gamma_{\min}^{(1)} \leq \gamma_{\min}^{(2)} \implies & \operatorname{EE}_{\textnormal{worst}}^\star (\delta,\gamma_{\min}^{(1)}) \geq \operatorname{EE}_{\textnormal{worst}}^\star (\delta,\gamma_{\min}^{(2)}) ,  \\
  & \; \forall \delta \in \mathcal{A}  ,
\end{split}
\end{equation}
where $\mathcal{A} = \{ \delta \geq 0 : \, \delta \ \textnormal{satisfies inequality}\ \eqref{equation:delta_psi} \}$.
\end{proposition}

\begin{IEEEproof}
First of all, Theorem \ref{theorem:Worst-case_SNR} is applicable because of Assumption \ref{assumption:CSI-uncertainty_radius_1}. Now, let $\mathcal{F}(\delta,\gamma_{\min}) = \{\mathbf{x} \in \{ 0,1 \}^L : \, {\gamma_{\text{worst}}} (\mathbf{x};\delta) \geq \gamma_{\min} \}$ denote the feasible set of problem \eqref{problem:EE_original}. Moreover, we have ${\partial g(\mathbf{x};\delta)} / {\partial \delta} = \sqrt{1 + \sum_{\ell \in \mathcal{L}} {x_\ell}} \geq 1 > 0 \implies {\partial {\gamma_{\text{worst}}} (\mathbf{x};\delta)} / {\partial \delta} = - 2\overline{\gamma} (f(\mathbf{x}) - g(\mathbf{x};\delta)) {{\partial g(\mathbf{x};\delta)} / {\partial \delta}} \leq 0 \implies {\partial {\operatorname{EE}_{\text{worst}}}(\mathbf{x};\delta)} / {\partial \delta} = [\log(2) {\operatorname{P}_\text{tot} (\mathbf{x})} (1 + {\gamma_{\text{worst}}} (\mathbf{x};\delta))]^{-1} {\partial {\gamma_{\text{worst}}} (\mathbf{x};\delta)} / {\partial \delta} \leq 0$, i.e., ${\gamma_{\text{worst}}} (\mathbf{x};\delta)$ and ${\operatorname{EE}_{\text{worst}}}(\mathbf{x};\delta)$ are nonincreasing functions of $\delta$. Consequently, we obtain the following implications: 1) $\delta_1 \leq \delta_2 \implies [{\gamma_{\text{worst}}} (\mathbf{x};\delta_1) \geq {\gamma_{\text{worst}}} (\mathbf{x};\delta_2)] \land [{\operatorname{EE}_{\text{worst}}}(\mathbf{x};\delta_1) \geq {\operatorname{EE}_{\text{worst}}}(\mathbf{x};\delta_2)] \land [\mathcal{F}(\delta_2,\gamma_{\min}) \subseteq \mathcal{F}(\delta_1,\gamma_{\min})] \implies \operatorname{EE}_{\textnormal{worst}}^\star (\delta_1,\gamma_{\min}) = \max_{\mathbf{x} \in \mathcal{F}(\delta_1,\gamma_{\min})} {{\operatorname{EE}_{\text{worst}}}(\mathbf{x};\delta_1)} \geq \max_{\mathbf{x} \in \mathcal{F}(\delta_2,\gamma_{\min})} {{\operatorname{EE}_{\text{worst}}}(\mathbf{x};\delta_1)} \geq \max_{\mathbf{x} \in \mathcal{F}(\delta_2,\gamma_{\min})} {{\operatorname{EE}_{\text{worst}}}(\mathbf{x};\delta_2)} = \operatorname{EE}_{\textnormal{worst}}^\star (\delta_2,\gamma_{\min})$, and 2) $\gamma_{\min}^{(1)} \leq \gamma_{\min}^{(2)} \implies \mathcal{F}(\delta,\gamma_{\min}^{(2)}) \subseteq \mathcal{F}(\delta,\gamma_{\min}^{(1)}) \implies \operatorname{EE}_{\textnormal{worst}}^\star (\delta,\gamma_{\min}^{(1)}) = \max_{\mathbf{x} \in \mathcal{F}(\delta,\gamma_{\min}^{(1)})} {{\operatorname{EE}_{\text{worst}}}(\mathbf{x};\delta)} \geq \max_{\mathbf{x} \in \mathcal{F}(\delta,\gamma_{\min}^{(2)})} {{\operatorname{EE}_{\text{worst}}}(\mathbf{x};\delta)} = \operatorname{EE}_{\textnormal{worst}}^\star (\delta,\gamma_{\min}^{(2)})$. 
\end{IEEEproof}

\section{Global Optimization Using Dynamic Programming} \label{section:Dynamic_Programming}

The discrete optimization problem \eqref{problem:EE_original} can be globally solved using \emph{exhaustive search}, however, with \emph{exponential complexity} $\Theta \left( \sum_{k=0}^L {\binom{L}{k} k} \right) = \Theta ({2^L}L)$. Therefore, exhaustive search is impractical for relatively large $L$, which is usually the case. Motivated by this fact, we will design a \emph{DP algorithm} that achieves the \emph{global optimum} in \emph{polynomial time} $O(L\,{\log L})$.\footnote{A DP algorithm was also designed in \cite{Efrem2021} for the selection of ground stations in satellite systems so as to minimize the total installation cost.} Thus, the problem is computationally tractable despite being nonconvex.

Subsequently, problem \eqref{problem:EE_original} can be decomposed into $L+1$ subproblems of the following form, where $M \in \mathcal{L}_0$, 
\begin{subequations}  \label{subproblem:equal_M}
\begin{alignat}{3}
  \operatorname{EE}^\star_{=M} \! \triangleq  & \mathop {\text{maximize}} \limits_{\mathbf{x} \in \{ 0,1 \}^L} & \quad & {\operatorname{EE}_\text{worst}}(\mathbf{x};\delta)  \\
  & \,\text{subject to} & & {\gamma_\text{worst}}(\mathbf{x};\delta) \geq \gamma_{\min} , \\
  & & & \sum_{\ell \in \mathcal{L}} {x_\ell} = M  .
\end{alignat}
\end{subequations}
Henceforth, we will write $\operatorname{EE}_{\text{worst}}^\star$ instead of $\operatorname{EE}_{\text{worst}}^\star (\delta,\gamma_{\min})$, and make the convention that if a maximization problem is not feasible, then its optimal value is set to $-\infty$. 

\begin{remark} \label{remark:Subproblems_decomposition}
The decomposition of \eqref{problem:EE_original} is essentially a partition of its feasible set, thus $\operatorname{EE}^\star_\textnormal{worst} = \max_{M \in \mathcal{L}_0} \{ \operatorname{EE}^\star_{=M} \}$. In addition, problem \eqref{problem:EE_original} is feasible if and only if there exists an $M \in \mathcal{L}_0$ such that subproblem \eqref{subproblem:equal_M} is feasible. 
\end{remark}

According to the following proposition, problem \eqref{subproblem:equal_M} can be solved by selecting $M$ IRS elements with the largest $\widehat{\alpha}_\ell$.

\begin{proposition} \label{proposition:Equivalent_subproblem}
Suppose that Assumption \ref{assumption:CSI-uncertainty_radius_1} is true. Then, for any given $M \in \mathcal{L}_0$, subproblem \eqref{subproblem:equal_M} is equivalent to\footnote{We say that two optimization problems are ``equivalent'' if they have equal sets of optimal solutions, i.e., any optimal solution to the first problem is also an optimal solution to the second problem and vice versa.}
\begin{subequations}  \label{subproblem:equal_M_equivalent}
\begin{alignat}{3}
 & \mathop {\textnormal{maximize}} \limits_{\mathbf{x} \in \{ 0,1 \}^L} & \quad & f(\mathbf{x}) \\
  & \,\textnormal{subject to} & & f(\mathbf{x}) \geq f_{\min}(M) , \\
  & & & \sum_{\ell \in \mathcal{L}} {x_\ell} = M  ,
\end{alignat}
\end{subequations}
where $f_{\min}(M) = \sqrt{{\gamma_{\min}}/{\overline{\gamma}}} + \delta \sqrt{1+M}$. Moreover, subproblem \eqref{subproblem:equal_M}/\eqref{subproblem:equal_M_equivalent} is feasible if and only if $\widehat{\alpha}_0 + \sum_{m=1}^{M} {\widehat{\alpha}_{\zeta_m}} \geq f_{\min}(M)$, where $(\zeta_1,\dots,\zeta_L)$ is a permutation of $\mathcal{L}$ such that $\widehat{\alpha}_{\zeta_1} \geq \cdots \geq \widehat{\alpha}_{\zeta_L}$. In addition, an optimal solution to subproblem \eqref{subproblem:equal_M}/\eqref{subproblem:equal_M_equivalent}, provided that it is feasible, is given by: $x_{\zeta_m}^\star = 1$, $\forall m \in \mathcal{M}=\{1,\dots,M\}$, and $x_{\zeta_\ell}^\star = 0$, $\forall \ell \in {\mathcal{L} \setminus \mathcal{M}}$. 
\end{proposition}

\begin{IEEEproof}
Based on Theorem \ref{theorem:Worst-case_SNR}, a key observation is that, for fixed $\sum_{\ell \in \mathcal{L}} {x_\ell} = M$, the worst-case SNR and total power consumption are written as $\gamma_\text{worst} (\mathbf{x};\delta) = \overline{\gamma} \left( f(\mathbf{x}) - \delta \sqrt{1+M} \right)^2$ and $\operatorname{P}_\text{tot} (\mathbf{x}) =  P_{\text{fix}} + L P_{\text{off}} + (P_{\text{on}} - P_{\text{off}}) M$. As a result, we obtain the problem equivalence, since $\log_2(y)$ and $y^2$ are both increasing functions for $y>0$. The if-and-only-if statement about feasibility and the optimal solution follow directly from the fact that $\max_{\mathbf{x} \in \{ 0,1 \}^L} \left\{ f(\mathbf{x}) : \, \sum_{\ell \in \mathcal{L}} {x_\ell} = M \right\} = \widehat{\alpha}_0 + \sum_{m=1}^{M} {\widehat{\alpha}_{\zeta_m}}$. 
\end{IEEEproof}

\noindent Also, we define the following subproblem, for any $M \in \mathcal{L}_0$, 
\begin{subequations}  \label{subproblem:less_equal_M}
\begin{alignat}{3}
  \operatorname{EE}^\star_{\leq M} \! \triangleq  & \mathop {\text{maximize}} \limits_{\mathbf{x} \in \{ 0,1 \}^L} & \quad & {\operatorname{EE}_\text{worst}}(\mathbf{x};\delta)  \\
  & \,\text{subject to} & & {\gamma_\text{worst}}(\mathbf{x};\delta) \geq \gamma_{\min} , \\
  & & & \sum_{\ell \in \mathcal{L}} {x_\ell} \leq M  .
\end{alignat}
\end{subequations}
This subproblem will facilitate the identification of the \linebreak \emph{optimal substructure}, which is a fundamental ingredient of DP \cite{Cormen2009}.\footnote{An optimization problem is said to have ``optimal substructure'' if its optimal value can be computed using the optimal values of its subproblems.} Since the constraint $\sum_{\ell \in \mathcal{L}} {x_\ell} \leq M$ is equivalent to $\left( \sum_{\ell \in \mathcal{L}} {x_\ell} \leq M-1 \right) \lor \left( \sum_{\ell \in \mathcal{L}} {x_\ell} = M \right)$, we have the following \emph{recurrence relation}  
\begin{equation} \label{equation:Recurrence_relation}
\operatorname{EE}_{\leq {M}}^\star = \max({\operatorname{EE}_{\leq M-1}^\star},{\operatorname{EE}_{=M}^\star}), \;\; \forall M \in \mathcal{L} ,
\end{equation}
with initial condition $\operatorname{EE}_{\leq 0}^\star = \operatorname{EE}_{=0}^\star$. 

\begin{remark} \label{remark:Recurrence_relation}
$\left\{ \operatorname{EE}_{=M}^\star \right\}_{M \in \mathcal{L}_0}$ can be computed by exploiting Proposition \ref{proposition:Equivalent_subproblem}. Moreover, $\operatorname{EE}_{\leq L}^\star = \operatorname{EE}^\star_\textnormal{worst}$ because the constraint $\sum_{\ell \in \mathcal{L}} {x_\ell} \leq L$ holds for all $\mathbf{x} \in \{0,1\}^L$, so it can be omitted. In other words, subproblem \eqref{subproblem:less_equal_M} for $M=L$ is identical to problem \eqref{problem:EE_original}. We can also prove, using mathematical induction, that the solution of \eqref{equation:Recurrence_relation} is $\operatorname{EE}^\star_\textnormal{worst} = \operatorname{EE}_{\leq L}^\star = \max_{M \in \mathcal{L}_0} \{ \operatorname{EE}^\star_{=M} \}$, which is in agreement with Remark \ref{remark:Subproblems_decomposition}. 
\end{remark}

The DP approach is presented in Algorithm \ref{algorithm:DP}. First, the algorithm sorts the entries of $[{\widehat{\alpha}_1}, \ldots ,{\widehat{\alpha}_L}]^\top$ in descending order and initializes some programming variables (steps 1--8). Subsequently, the for-loop (steps 9--15) updates the values of $\operatorname{EE}^\star$ and $M^\star$ according to the recurrence relation \eqref{equation:Recurrence_relation}. Finally, steps 16--24 decide the feasibility of problem \eqref{problem:EE_original}, and reconstruct an optimal solution if the problem is feasible. More precisely, the correctness and polynomial complexity of the proposed algorithm are established by the following theorem.

\begin{theorem} \label{theorem:DP_algorithm}
Provided that Assumption \ref{assumption:CSI-uncertainty_radius_1} holds, Algorithm \ref{algorithm:DP} returns either \textit{``Infeasible''} if problem \eqref{problem:EE_original} is not feasible, or its global maximum together with an optimal solution otherwise. Furthermore, the complexity of Algorithm \ref{algorithm:DP} is $O(L\,{\log L})$.
\end{theorem}

\begin{IEEEproof}
See Appendix \ref{appendix:B}.
\end{IEEEproof}

\noindent
\begin{minipage}[!t]{\columnwidth}
\begin{algorithm}[H]   
\caption{Dynamic Programming}  \label{algorithm:DP}
\small
\begin{algorithmic}[1] 
\State Sort the entries of $[{\widehat{\alpha}_1}, \ldots ,{\widehat{\alpha}_L}]^\top$ in descending order. Let $({\zeta_1}, \ldots ,{\zeta_L})$ be a permutation of $\mathcal{L}$ such that $\widehat{\alpha}_{\zeta_1} \geq \cdots \geq \widehat{\alpha}_{\zeta_L}$.
\State $f :=  \widehat{\alpha}_0$, $\gamma_\text{worst} := \overline{\gamma} \left( f - \delta \right)^2$
\State $\operatorname{P}_\text{tot} :=  P_{\text{fix}} + L P_{\text{off}}$, $\operatorname{\Delta P} := P_{\text{on}} - P_{\text{off}}$ 
\If{$\gamma_\text{worst} \geq \gamma_{\min}$}
	\State  ${\operatorname{EE}^\star} := \log_2(1+\gamma_\text{worst})/\operatorname{P}_\text{tot}$, $M^\star := 0$ 
\Else 
	\State ${\operatorname{EE}^\star} := -\infty$ 
\EndIf 
\For{$M : = 1\ \text{to}\ L$}
	\State $f := f + \widehat{\alpha}_{\zeta_M}$, $\gamma_\text{worst} := \overline{\gamma} \left( f - \delta \sqrt{1+M} \right)^2$
	\State $\operatorname{P}_\text{tot} :=  \operatorname{P}_\text{tot} + \operatorname{\Delta P}$, $\operatorname{EE} := \log_2(1+\gamma_\text{worst})/\operatorname{P}_\text{tot}$ 
	\If{$(\gamma_\text{worst} \geq \gamma_{\min}) \land (\operatorname{EE} > \operatorname{EE}^\star)$}
		\State $\operatorname{EE}^\star := \operatorname{EE}$, $M^\star := M$    
	\EndIf	
\EndFor 
\If{${\operatorname{EE}^\star} = -\infty$} 
	\State \textbf{return} \textit{``Infeasible''} 
\Else
	\State $\mathbf{x}^\star := \mathbf{0}_L$
	\For{$m := 1\ \text{to}\ M^\star$}
		\State $x_{\zeta_m}^\star := 1$
	\EndFor
	\State \textbf{return} $({\operatorname{EE}^\star},{\mathbf{x}^\star})$		
\EndIf
\end{algorithmic}
\end{algorithm} 
\end{minipage}

\section{Numerical Results} \label{section:Numerical_Results}

In this section, we evaluate the performance of the proposed algorithm via simulations, where the transmitter, receiver and IRS are located at $(0,0,0)$, $(100,0,0)$ and $(50,20,10)$, respectively (with $(x,y,z)$-coordinates given in meters). All figures present average values derived from $10^3$ independent random realizations of channel estimates $\widehat{\mathbf{h}}$. In particular, $\widehat{h}_0 \sim \mathcal{CN}(0,\varrho_0)$, where $\varrho_0 = c_0^\text{ref} ({d_0}/{d_0^\text{ref}})^{-a_0}$ is the distance-dependent path loss, with $d_0$, $a_0$ being the Tx-Rx distance and path-loss exponent, and $d_0^\text{ref}$, $c_0^\text{ref}$ being the reference distance and path loss, respectively. Moreover, $\widehat{h}_\ell = {\beta_\ell} {\widehat{u}_\ell} {\widehat{v}_\ell}$, $\forall \ell \in \mathcal{L}$, where $\widehat{u}_\ell = \sqrt{\varrho_u} \left( \sqrt{\frac{\kappa_u}{1+\kappa_u}} {\widehat{u}_\ell^\text{LOS}} + \sqrt{\frac{1}{1+\kappa_u}} {\widehat{u}_\ell^\text{NLOS}} \right)$ and $\widehat{v}_\ell =  \sqrt{\varrho_v} \left( \sqrt{\frac{\kappa_v}{1+\kappa_v}} {\widehat{v}_\ell^\text{LOS}} + \sqrt{\frac{1}{1+\kappa_v}} {\widehat{v}_\ell^\text{NLOS}} \right)$, with $\kappa_{u/v}$ being the Rician factors and $\varrho_{u/v} = c_{u/v}^\text{ref} ({d_{u/v}}/{d_{u/v}^\text{ref}})^{-a_{u/v}}$; the subscripts $u$ and $v$ refer to Tx-IRS and IRS-Rx links, respectively. Assuming a uniform linear array of reflecting elements at the IRS (parallel to the $x$-axis), the line-of-sight (LOS) components are given by $\widehat{u}_\ell^\text{LOS} = e^{j2\pi \tfrac{d}{\lambda} (\ell - 1) \sin(\vartheta^\text{A}) \cos(\varphi^\text{A})}$ and $\widehat{v}_\ell^\text{LOS} = e^{j2\pi \tfrac{d}{\lambda} (\ell - 1) \sin(\vartheta^\text{D}) \cos(\varphi^\text{D})}$, where $d$ is the distance between adjacent IRS elements, and $\lambda$ is the wavelength of the carrier frequency. In addition, $\vartheta^\text{A}$ and $\varphi^\text{A}$ are the inclination and azimuth angle-of-arrival (AoA) of signals from the Tx to the IRS, respectively. Similarly, $\vartheta^\text{D}$ and $\varphi^\text{D}$ are the inclination and azimuth angle-of-departure (AoD) of signals from the IRS to the Rx, respectively. The non-line-of-sight (NLOS) components ${\widehat{u}_\ell^\text{NLOS}}, {\widehat{v}_\ell^\text{NLOS}} \sim \mathcal{CN}(0,1)$. The simulation parameters are $c_0^\text{ref} = 10^{-5}$, $c_u^\text{ref} = c_v^\text{ref} = 10^{-3}$, $d_0^\text{ref} = d_u^\text{ref} = d_v^\text{ref} = 1\ \text{m}$, $a_0 = 3.7$, $a_u = a_v = 2.2$, $\kappa_u = \kappa_v = 5\ \text{dB}$, and $d/\lambda = 0.5$.  

The CSI-uncertainty radius is selected as $\delta = \tau {\widehat{\alpha}_{\min}}$, where $\tau \in [0,1]$, so that $\delta \leq {\widehat{\alpha}_{\min}}$, thus satisfying Assumption \ref{assumption:CSI-uncertainty_radius_2} (and Assumption \ref{assumption:CSI-uncertainty_radius_1} as well, according to Proposition \ref{proposition:Assumptions_1_2}). The minimum SNR is chosen to be $\gamma_{\min} = {\nu} {\gamma}_{\text{worst}} (\mathbf{1}_L;\widehat{\alpha}_{\min})$, where $\nu \in [0,1]$ is the minimum-SNR control parameter. In this way, the feasibility of generated optimization problems is guaranteed, because ${\gamma}_{\text{worst}} (\mathbf{1}_L;\delta) \geq {\gamma}_{\text{worst}} (\mathbf{1}_L;\widehat{\alpha}_{\min}) \geq \gamma_{\min}$ (see Propositions \ref{proposition:Feasibility} and \ref{proposition:Optimal_worst-case_EE_monotonicity}). Unless otherwise stated, the remaining simulation parameters are the following: $L = 20$, $p = 10\ \text{dBm}$, $\sigma_w^2 = -120\ \text{dBm}$, $\beta_\ell = 0.9,\ \forall \ell \in \mathcal{L}$, $\eta = 0.8$, $P_{\text{static}} = 10\ \text{mW}$, $P_\text{on} = 1.5\ \text{mW}$, $P_\text{off} = 0.3\ \text{mW}$, and $\nu = 0.7$. Besides the exhaustive-search method, we also consider a baseline scheme, namely, the activation of all IRS elements (i.e., $\mathbf{x} = \mathbf{1}_L$).\footnote{Note that $\mathbf{x} = \mathbf{1}_L$ is an optimal solution to the worst-case SE maximization (under the minimum-SNR constraint). This is because Assumption \ref{assumption:CSI-uncertainty_radius_2} is true in all simulation scenarios, and ${\gamma}_{\textnormal{worst}} (\mathbf{1}_L;\delta) \geq {\gamma_{\text{worst}}} (\mathbf{x};\delta)$, $\forall \mathbf{x} \in \{0,1\}^L$, due to Proposition \ref{proposition:Feasibility}.}

\begin{figure}[!t]
\centering
\includegraphics[width=\columnwidth]{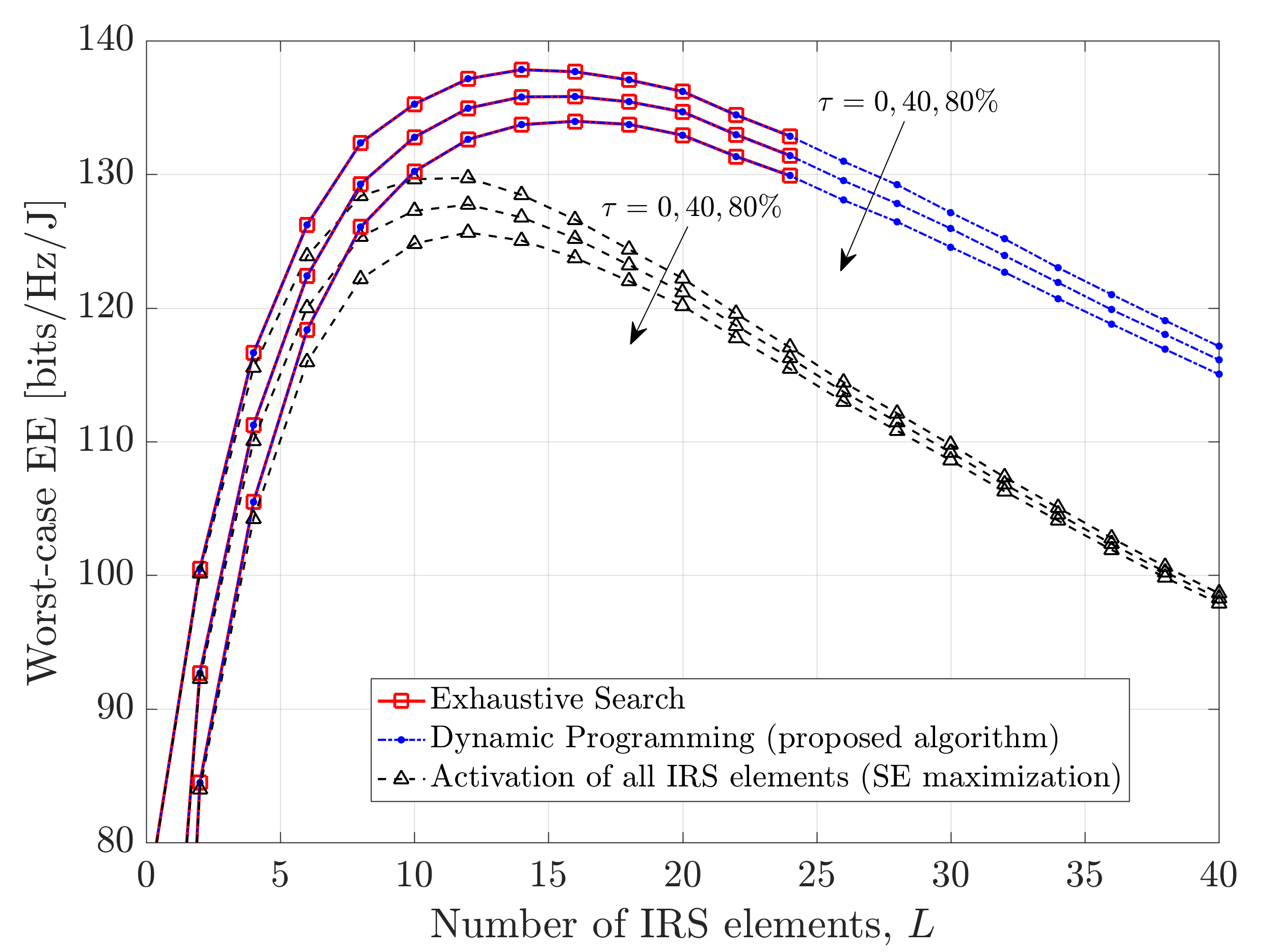}
\caption{Worst-case energy efficiency versus the number of IRS elements, for several CSI-uncertainty radii ($\delta = \tau {\widehat{\alpha}_{\min}}$).}
\label{figure:EE_vs_L}
\end{figure}

First of all, Fig. \ref{figure:EE_vs_L} shows the worst-case EE versus the number of IRS elements.\footnote{We have run the exhaustive-search algorithm only for $L \leq 24$, because of its exponential complexity.} The proposed algorithm has identical performance with the exhaustive search (as expected by Theorem \ref{theorem:DP_algorithm}), while performing much better than the baseline scheme for moderate to large $L$. For small $L$, however, all algorithms achieve roughly the same worst-case EE; this is because $P_{\text{fix}} + L P_{\text{off}} \gg (P_{\text{on}} - P_{\text{off}}) {\sum_{\ell \in \mathcal{L}} {x_\ell}} \implies \operatorname{P}_\text{tot} (\mathbf{x}) \approx P_{\text{fix}} + L P_{\text{off}}, \ \forall \mathbf{x} \in \{0,1\}^L$, and therefore EE maximization approximately reduces to SE maximization. In addition, for all algorithms, the worst-case EE decreases as the CSI-uncertainty radius increases, which is in agreement with Proposition \ref{proposition:Optimal_worst-case_EE_monotonicity} (note that $\gamma_{\min}$ is independent of $\delta$). Consequently, the highest worst-case EE occurs in the perfect-CSI regime, i.e., for $\delta = 0$. It is also interesting to observe that the worst-case EE increases rapidly up to a point, and then decreases slowly (since additional IRS elements begin to consume a significant amount of power, even if they are off/deactivated). In particular, the exhaustive search and DP algorithm reach their peak for some $L \in \{14,15,16\}$, whereas the baseline scheme for some smaller $L \in \{10,11,12\}$.

\begin{figure}[!t]
\centering
\includegraphics[width=\columnwidth]{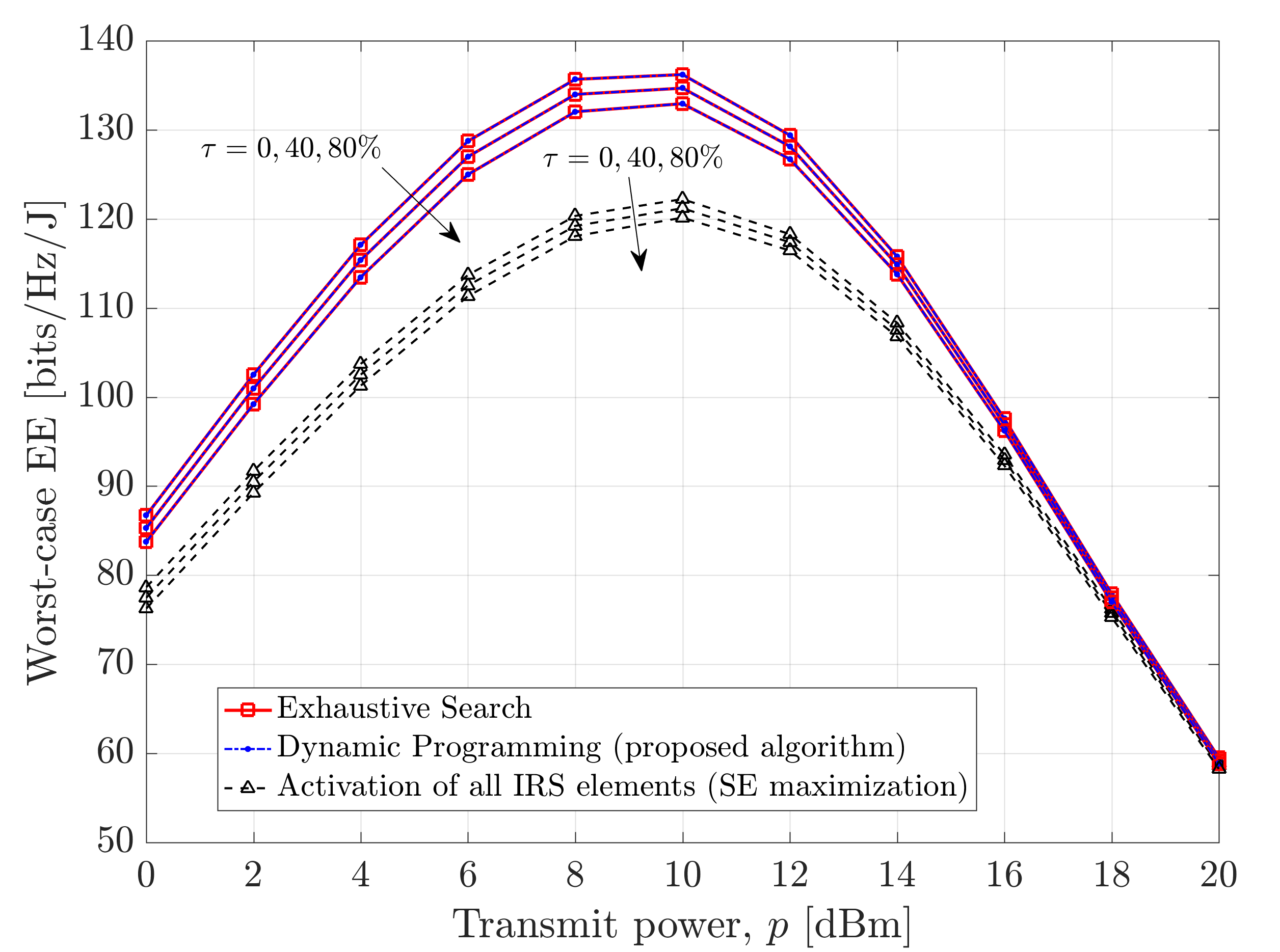}
\caption{Worst-case energy efficiency versus the transmit power, for different CSI-uncertainty radii ($\delta = \tau {\widehat{\alpha}_{\min}}$).}
\label{figure:EE_vs_p}
\end{figure}

Moreover, Fig. \ref{figure:EE_vs_p} presents the worst-case EE against the transmit power. Again, the worst-case EE of DP algorithm coincides with that of the exhaustive search, and is higher than that of the benchmark. We can also observe the monotonicity of the worst-case EE with respect to the CSI-uncertainty radius, in line with Proposition \ref{proposition:Optimal_worst-case_EE_monotonicity}. All algorithms attain their maximum for $p = 10\ \text{dBm}$, while they show similar performance for high transmit power (in this case, $\operatorname{P}_\text{tot} (\mathbf{x}) \approx P_{\text{fix}} + L P_{\text{off}} \approx p/\eta, \ \forall \mathbf{x} \in \{0,1\}^L$, thus EE maximization tends to be equivalent to SE maximization as $p \to \infty$). 

\begin{figure}[!t]
\centering
\includegraphics[width=\columnwidth]{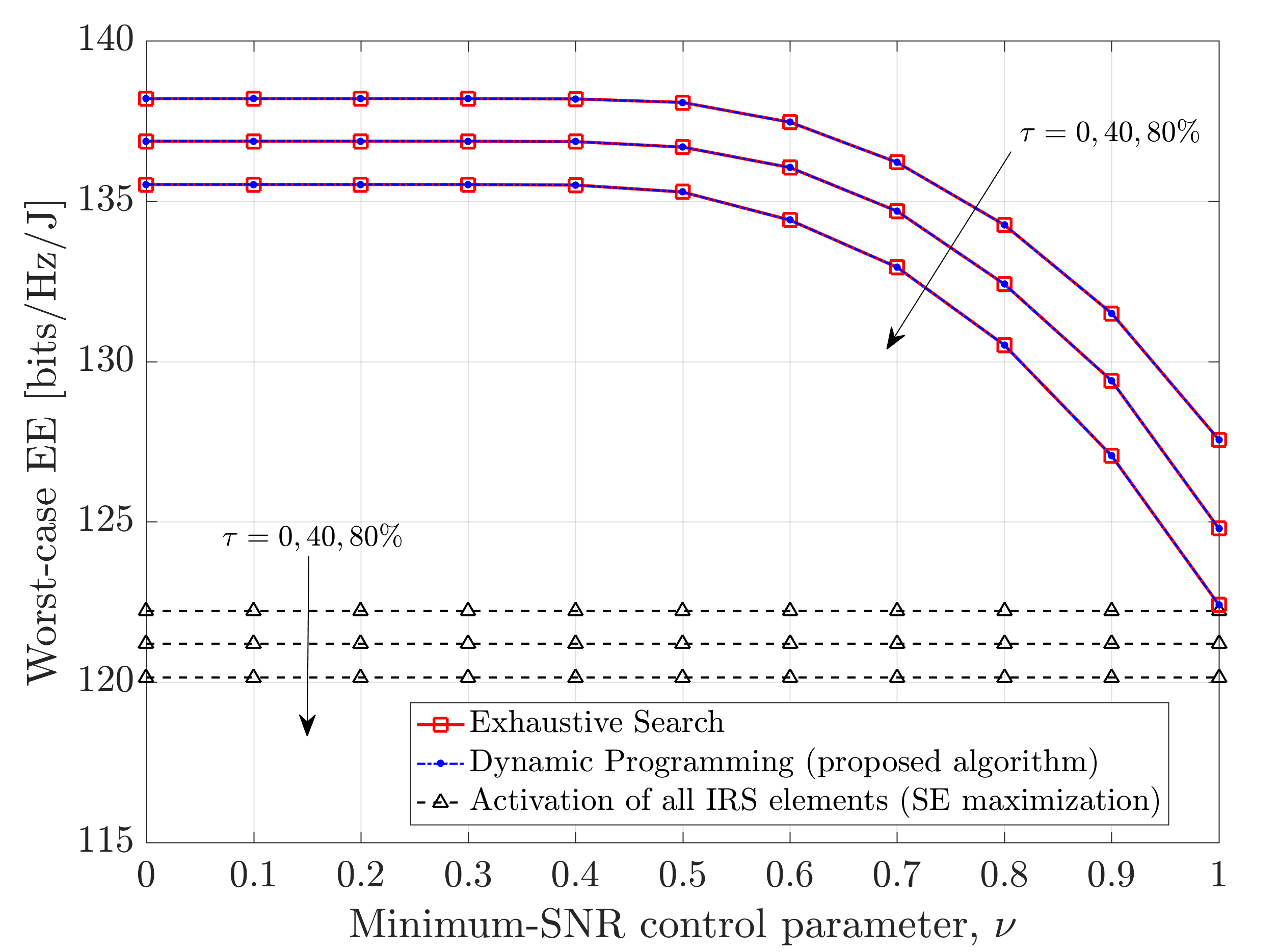}
\caption{Worst-case energy efficiency versus the minimum-SNR control parameter, for various CSI-uncertainty radii ($\delta = \tau {\widehat{\alpha}_{\min}}$).}
\label{figure:EE_vs_nu}
\end{figure}

Finally, we examine the impact of the minimum required SNR on the worst-case EE. Based on Fig. \ref{figure:EE_vs_nu}, we can make similar observations as in the previous figures. Furthermore, the optimal worst-case EE achieved by the proposed algorithm and exhaustive search is a nonincreasing function of $\nu$, since the increase of $\nu$ (or $\gamma_{\min}$) results in the shrinkage/contraction of the feasible set of problem \eqref{problem:EE_original}; see also Proposition \ref{proposition:Optimal_worst-case_EE_monotonicity}.

\section{Conclusion and Future Work} \label{section:Conclusion}

In this article, we dealt with the maximization of the worst-case EE in IRS-aided communication systems with imperfect CSI, via robust activation of IRS elements. Furthermore, we developed a polynomial-time algorithm with global optimization guarantee, based on dynamic programming. Numerical results verified the effectiveness of the proposed algorithm and demonstrated its superior performance compared to a conventional scheme. Finally, the impact of discrete IRS phase shifts (i.e., with finite bit-resolution) on the system performance is an interesting direction for future research.

\appendices

\section{Proof of Theorem \ref{theorem:Worst-case_SNR}} \label{appendix:A}

Firstly, observe that constraint \eqref{constraint:CSI_uncertainty} does not impose any restrictions on the principal arguments of $\widetilde{\mathbf{h}}$, apart from being in the interval $[0,2\pi)$, but only on its amplitudes. Consequently, by inspection of \eqref{equation:SNR_optimal_phase_shifts}, we can deduce that the minimum SNR is achieved when  
\begin{equation}
\widetilde{\theta}_0^\star = (\widehat{\theta}_0 + \pi) \bmod {2\pi}  ,
\end{equation}
\begin{equation}
\widetilde{\theta}_\ell^\star = (\widehat{\theta}_0 - {\phi}_\ell^{\star} + \pi) \bmod {2\pi} ,  \;\;\forall \ell \in \mathcal{L} .
\end{equation}
In other words, the arguments of $\widetilde{\mathbf{h}}$ are adjusted so as to cause the largest possible decrease in $f(\mathbf{x})$; in particular, $\widetilde{h}_0 + \sum_{\ell \in \mathcal{L}} {x_\ell \widetilde{h}_\ell e^{j{\phi}_\ell^{\star}}}$ is aligned with $f(\mathbf{x}) e^{j\widehat{\theta}_0}$, but in the opposite direction. 

Hence, problem \eqref{problem:Worst-case_SNR} is equivalent to the following problem 
\begin{subequations} \label{problem:Absolute_value}
\begin{alignat}{3}
  & \mathop {\text{minimize}} \limits_{\widetilde{\boldsymbol{\alpha}} \in \mathbb{R}_{+}^{L+1}} & \quad &  \left| f(\mathbf{x}) - \left( \widetilde{\alpha}_0 + \sum_{\ell \in \mathcal{L}} {x_\ell \widetilde{\alpha}_\ell} \right) \right|   \\
  & \,\text{subject to} & & \sqrt{ \widetilde{\alpha}_0^2 + \sum_{\ell \in \mathcal{L}} {\widetilde{\alpha}_\ell^2} } \leq \delta ,
\end{alignat}
\end{subequations}
where $\widetilde{\boldsymbol{\alpha}} = [{\widetilde{\alpha}_0}, \ldots ,{\widetilde{\alpha}_L}]^\top$. In order to facilitate the solution of problem \eqref{problem:Absolute_value}, we exploit the following lemma.

\begin{lemma}
Let $g(\mathbf{x};\delta)$ be the global maximum of the following optimization problem
\begin{subequations}  \label{problem:g}
\begin{alignat}{3}
 g(\mathbf{x};\delta) \! \triangleq & \mathop {\textnormal{maximize}} \limits_{\widetilde{\boldsymbol{\alpha}} \in \mathbb{R}_{+}^{L+1}} & \quad & \widetilde{\alpha}_0 + \sum_{\ell \in \mathcal{L}} {x_\ell \widetilde{\alpha}_\ell}   \\
  & \,\textnormal{subject to} & & \sqrt{ \widetilde{\alpha}_0^2 + \sum_{\ell \in \mathcal{L}} {\widetilde{\alpha}_\ell^2} } \leq \delta  .
\end{alignat}
\end{subequations}
Then, $g(\mathbf{x};\delta) = \delta \sqrt{1 + \sum_{\ell \in \mathcal{L}} {x_\ell}} \,$, which is achieved for $\widetilde{\alpha}_0^\star = \widetilde{\alpha}_\ell^\star = \delta \big/ \sqrt{1 + \sum_{l \in \mathcal{L}} {x_l}} \,$,  $\forall \ell \in \mathcal{P}$, and $\widetilde{\alpha}_\ell^\star = 0$,  $\forall \ell \in {\mathcal{L} \setminus \mathcal{P}}$, where $\mathcal{P} = \{\ell \in \mathcal{L} : \, x_\ell = 1 \}$.
\end{lemma}

\begin{IEEEproof}
Obviously, $|\mathcal{P}| = \sum_{\ell \in \mathcal{L}} {x_\ell}$ and $\sum_{\ell \in \mathcal{L}} {x_\ell \widetilde{\alpha}_\ell} = \sum_{\ell \in \mathcal{P}} {\widetilde{\alpha}_\ell}$. Now, we examine the following relaxation problem 
\begin{subequations} \label{problem:g_rel}
\begin{alignat}{3}
 g_\text{rel}(\mathbf{x};\delta) \! \triangleq & \mathop {\text{maximize}} \limits_{\widetilde{\boldsymbol{\alpha}}_{\mathcal{P}_0} \in \mathbb{R}_{+}^{|\mathcal{P}|+1}} & \quad & \widetilde{\alpha}_0 + \sum_{\ell \in \mathcal{P}} {\widetilde{\alpha}_\ell}   \\ 
  & \,\text{subject to} & & \sqrt{ \widetilde{\alpha}_0^2 + \sum_{\ell \in \mathcal{P}} {\widetilde{\alpha}_\ell^2} } \leq \delta ,
\end{alignat}
\end{subequations}
where $\widetilde{\boldsymbol{\alpha}}_{\mathcal{P}_0} = [{\widetilde{\alpha}_\ell}]_{\ell \in \mathcal{P}_0}^\top$, with ${\mathcal{P}_0} = \{0\} \cup \mathcal{P}$. Since $\sqrt{ \widetilde{\alpha}_0^2 + \sum_{\ell \in \mathcal{P}} {\widetilde{\alpha}_\ell^2} } \leq \sqrt{ \widetilde{\alpha}_0^2 + \sum_{\ell \in \mathcal{L}} {\widetilde{\alpha}_\ell^2} }$, we have that $g(\mathbf{x};\delta) \leq g_\text{rel}(\mathbf{x};\delta)$. On the other hand, an optimal solution of the relaxation problem \eqref{problem:g_rel} together with $\widetilde{\alpha}_\ell = 0$, $\forall \ell \in {\mathcal{L} \setminus \mathcal{P}}$, constitute a feasible solution of problem \eqref{problem:g}, so $g(\mathbf{x};\delta) \geq g_\text{rel}(\mathbf{x};\delta)$. By combining both results, we obtain $g(\mathbf{x};\delta) = g_\text{rel}(\mathbf{x};\delta)$. 

Subsequently, in order to solve problem \eqref{problem:g_rel}, we make use of the \emph{Cauchy-Bunyakovsky-Schwarz  inequality}: For any $\mathbf{y},\mathbf{z} \in \mathbb{R}^N$, it holds that
\begin{equation}
\left| \sum_{n=1}^N {y_n z_n} \right| \leq \sqrt{\sum_{n=1}^N {y_n^2}} \sqrt{\sum_{n=1}^N {z_n^2}} \, .
\end{equation} 
The above inequality holds with equality if and only if (iff) $\mathbf{y} = \lambda \mathbf{z}$ for some $\lambda \in \mathbb{R}$. Specifically, by setting $\mathbf{z} = \mathbf{1}_N$, we obtain $\left| \sum_{n=1}^N {y_n} \right| \leq \sqrt{N} \sqrt{\sum_{n=1}^N {y_n^2}}$ with equality iff $y_1=\cdots=y_N=\lambda$. Consequently, we obtain  
\begin{equation}
\widetilde{\alpha}_0 + \sum_{\ell \in \mathcal{P}} {\widetilde{\alpha}_\ell} \leq \sqrt{1 + |\mathcal{P}|} \sqrt{ \widetilde{\alpha}_0^2 + \sum_{\ell \in \mathcal{P}} {\widetilde{\alpha}_\ell^2} } \leq \delta \sqrt{1 + |\mathcal{P}|} \, ,
\end{equation}
with equalities (i.e., achieving the maximum value) when $\widetilde{\alpha}_0 = \widetilde{\alpha}_\ell = \lambda \geq 0$, $\forall \ell \in \mathcal{P}$, where $\sqrt{\lambda^2 (1+|\mathcal{P}|)} = \delta \implies \lambda = \delta \big/ \sqrt{1+|\mathcal{P}|}$. Thus, $g(\mathbf{x};\delta) = g_\text{rel}(\mathbf{x};\delta) = \delta \sqrt{1 + |\mathcal{P}|}$.
\end{IEEEproof}

Next, by leveraging Assumption \ref{assumption:CSI-uncertainty_radius_1}, we can easily prove the following equivalences 
\begin{equation} 
\begin{split}
 {\delta \leq \psi} & \iff {\widehat{\alpha}_0 + \sum_{m=1}^{M} {\widehat{\alpha}_{\sigma_m}} \geq \delta \sqrt{1 + M}, \;\; \forall M \in \mathcal{L}_0}  \\
 & \iff {f(\mathbf{x}) \geq g(\mathbf{x};\delta), \;\; \forall \mathbf{x} \in \{0,1\}^L}  .
\end{split}
\end{equation}
As a result, $f(\mathbf{x}) - \left( \widetilde{\alpha}_0 + \sum_{\ell \in \mathcal{L}} {x_\ell \widetilde{\alpha}_\ell} \right) \geq f(\mathbf{x}) - g(\mathbf{x};\delta) \geq 0$, $\forall \mathbf{x} \in \{0,1\}^L$, $\forall \widetilde{\boldsymbol{\alpha}} \in \mathcal{U}_{\delta}$, where $\mathcal{U}_{\delta} = \left\{ \widetilde{\boldsymbol{\alpha}} \in \mathbb{R}_{+}^{L+1} : \, \sqrt{ \sum_{\ell \in \mathcal{L}_0} {\widetilde{\alpha}_\ell^2} } \leq \delta \right\}$. Therefore, problems \eqref{problem:Absolute_value} and \eqref{problem:g} are equivalent, and the worst-case SNR is given by \eqref{equation:Worst-case_SNR}. This completes the proof of Theorem \ref{theorem:Worst-case_SNR}.

\section{Proof of Theorem \ref{theorem:DP_algorithm}}  \label{appendix:B}

In order to prove the correctness of the algorithm, we will use a \emph{loop invariant} \cite{Cormen2009}, namely, 

\vspace{0.7mm}
\noindent $I(M)$: \emph{``At the end of iteration $M$ of the for-loop in steps 9--15, ${\operatorname{EE}^\star} = \operatorname{EE}^\star_{\leq M}$. In addition, if subproblem \eqref{subproblem:less_equal_M} is feasible, then an optimal solution to this subproblem is given by: $x_{\zeta_m}^\star = 1$, $\forall m \in {\mathcal{M}^\star}=\{1,\dots,M^\star\}$, and $x_{\zeta_\ell}^\star = 0$, $\forall \ell \in {\mathcal{L} \setminus {\mathcal{M}^\star}}$.''}
\vspace{0.7mm}

Now, we will show that $I(M)$ is true for all $M \in \mathcal{L}_0$, using mathematical induction. In particular, we can obtain the following: 1) steps 2--8 together with Proposition \ref{proposition:Equivalent_subproblem} imply that $I(0)$ is true before the first iteration of the for-loop (\emph{basis}), and 2) if $I(M-1)$ is true then, by virtue of Proposition \ref{proposition:Equivalent_subproblem} and recurrence relation \eqref{equation:Recurrence_relation}, $I(M)$ is true as well (\emph{inductive step}). Hence, the above argument has been proved.  

After the termination of the for-loop, $I(L)$ will be true. As a result, ${\operatorname{EE}^\star} = \operatorname{EE}^\star_{\leq L}$, but $\operatorname{EE}^\star_{\leq L} = \operatorname{EE}^\star_\text{worst}$ (see Remark \ref{remark:Recurrence_relation}), thus ${\operatorname{EE}^\star} = \operatorname{EE}^\star_\text{worst}$. If problem \eqref{problem:EE_original} is not feasible (i.e., $\operatorname{EE}^\star_\text{worst} = -\infty$), then Algorithm \ref{algorithm:DP} correctly returns \textit{``Infeasible''} (steps 16--17). Otherwise, if the problem is feasible (i.e., $\operatorname{EE}^\star_\text{worst} > -\infty$), then $I(L)$ implies that an optimal solution to problem \eqref{problem:EE_original} can be derived using $M^\star$.\footnote{Recall that subproblem \eqref{subproblem:less_equal_M} for $M=L$ coincides with problem \eqref{problem:EE_original}, according to Remark \ref{remark:Recurrence_relation}.} This optimal solution is correctly reconstructed and returned by the algorithm (steps 18--24).

Finally, regarding the complexity of Algorithm \ref{algorithm:DP}, the sorting procedure in step 1 requires $O(L\,{\log L})$ comparisons, while all the remaining steps take $O(L)$ time. Consequently, the overall complexity is $O(L\,{\log L} + L) = O(L\,{\log L})$.

\end{document}